# Dark energy enveloping a massive body

## Mark Israelit[1]

In the spatially closed homogeneous and isotropic universe filled with luminous matter, dark matter and dark energy a massive cosmic body is considered. It is shown that around this body the dark energy forms up a ball-like concentration, that reaches the horizon. The energy mass density, the pressure and mass of this dark energy ball are negative. It is very plausible that such a dark energy ball is causing and supporting the present cosmic acceleration.

The work is carried out in the framework of a Weyl-Dirac cosmology, the latter being developed in a previous work: *Gen. Rel. Grav*. **43,** 751 (2011).

### 1. Introduction

In the nowadays cosmology, the problems of dark matter and dark energy are in the spotlight of discussions and research [1-6].

Dark matter (DM) was predicted by Fritz Zwicky [7] already in 1933 and became very popular in the last 30 years. However the origin and nature of DM is under discussion still.

Dark energy (DE) was introduced in order to give an explanation to a phenomenon that was discovered at the end of the twenties century, the present cosmic acceleration [8-11] of our expanding universe. DE is now widely discussed, however its origin and nature are unknown still.

In the last years were made many attempts [5, 6] to derive DM and DE from quantum gravity, from string theories, or by introduction into the cosmological framework DM and DE functions having no physical, cosmological origin. In the review [5] a detailed list of relevant papers is presented. But up today there are no general accepted answers.

Let us go to the Weyl-Dirac enlargement of Einstein's general relativity. After Einstein [12] developed his general theory of relativity, in which gravitation is

Department of Physics and Mathematics, University of Haifa – Oranim 36006 ISRAEL.
E-mail: israelit@macam.ac.il

described in terms of geometry, Weyl [13] proposed a more general theory, in which electromagnetism is also described geometrically. However, this theory had some unsatisfactory features and did not gain general acceptance. Later Dirac [14] returned to Weyl's theory but introduced modifications which removed the earlier difficulties. Nathan Rosen [15] in discussing the Weyl-Dirac (W-D) theory pointed out that Dirac had chosen a particular value for a parameter appearing in the W-D variational principle and that this parameter could be taken differently. Rosen [15] showed that the parameter could be chosen so that, instead of an electromagnetic field, one gets a Proca [16] vector field, which may be interpreted as an ensemble of particles having finite mass and spin 1. It was suggested that these particles named weylons might constitute most of the dark matter [17]. The W-D framework [13,14,15,18,19] contains three geometric quantities: the metric tensor $g_{\mu\nu} = g_{\nu\mu}$, the Weyl connection vector $w^\mu$ and the Dirac gauge function $\beta$.

Recently, in the framework of the Weyl-Dirac theory, a spatially closed cosmological model was proposed [20]. It was assumed that the space-time of the universe has a chaotic Weylian microstructure but is described on a large scale by Riemannian geometry enriched by the Dirac gauge function $\beta$.

In that model, locally fields of the Weyl connection vector $w^\mu$ act as creators of massive bosons having spin 1, weylons. On a large scale the universe was considered as a homogeneous and isotropic space, containing conventional (luminous) matter as well weylon dark matter. The space-time is penetrated by the scalar field of the Dirac gauge function $\beta$. The latter being source of dark energy that in the very early universe create conventional matter, whereas at present the Weylian DE causes the global acceleration of the expanding universe.

**In the present work** the changes caused by a massive cosmic body located in the homogeneous W-D universe are considered. The metric as well all functions describing the cosmic substances now depend on time and on a radial coordinate. The explicit equations are derived and solved. Around the massive body the $\beta-\text{DE}$ forms a spherically symmetric ball-like concentration that arrives the cosmic horizon, $r=1$. The energy density, the pressure as well the mass of the DE ball turn out to be negative. The negative gravitational mass is universally *repulsive*, both positive-mass and negative-mass objects will be pushed away. It is plausible to think that features of DE balls play an important role in causing and supporting the acceleration of our expanding universe.

In section **2** we give a concise review of the formalism developed in [20] for the homogeneous and isotropic Weyl-Dirac universe, that serves as background for the present model. Section **3** contains the general formalism of the model with the massive body. In section **4** the field equations are derived explicitly and the metric is obtained. It is an a' la Reissner-Nordstrøm [21], [22] metric, with two horizons, the inner and outer one. In section **5** the dark energy in the gravitational field of the massive body *m* is accounted and interpreted. This $\beta-\text{DE}$ forms a ball around *m*; at distances $2m < r \leq 1$ the mass-energy density, the pressure as well the mass of this $\beta-\text{DE}$ ball, all are negative. These features support the present cosmic acceleration. Section **6** contains an outlook and discussion.

**The following notations** are used below: Greek indices run from 0 to 3; a comma stands for the partial derivative, $f_{,\mu} \equiv \frac{\partial f}{\partial x^\mu}$; a semicolon stands for the covariant derivative formed with Christoffel indices, $V^\lambda_{;\mu} = V^\lambda_{,\mu} + V^\sigma \{^\lambda_{\sigma\mu}\}$ ; a dot denotes partial differentiation with respect to time, $\dot{f} \equiv \frac{\partial f}{\partial t} \lor \dot{f} \equiv \frac{\partial f}{\partial \tau}$ ; primes stand for partial differentiation with respect to the radial coordinate $r$, $f' \equiv \frac{\partial f}{\partial r}$. Further a tilde over the function, $\tilde{V}(t,r)$, indicates that it depends on both, $t$ and $r$; an over-bar, $\bar{V}(t)$, is used to denote functions depending only on $t$; quantities depending only on $r$ are not marked, $V(r)$. Indices may be lowered / raised with the metric tensor $g_{\mu\nu}$, $g^{\mu\nu}$. The following abbreviations are used throughout the present paper: W-D=Weyl- Dirac; DM=Weyl-Dirac Dark Matter consisting of weylons; DE=Dark Energy created by Dirac's gauge function.

### 1. The homogeneous and isotropic W-D universe (Cf. [20])

Let us begin with a brief review of the homogeneous and isotropic W-D universe, [20], that serves as a background for the present model. In order to obtain a cosmological model providing geometrically based Dark Matter (DM) and Dark Energy (DE) the W-D theory [13, 14, 18] was enriched by Nathan Rosen's approach [15, 17]. So we had a framework with three geometric quantities: the metric tensor $g_{\mu\nu} = g_{\nu\mu}$, the Weyl connection vector $w^\mu$ and the Dirac gauge function $\beta$. It is assumed that there exist chaotic oriented locally fields of the Weyl connection vector $w^\mu$ in microscopic cells, which act as creators of massive bosons having spin 1, weylons. Due to the chaotic orientation of the locally fields and spins only the mass effect interpreted as DM is recognized.

On a large scale the universe is homogeneous and isotropic containing conventional (luminous) matter as well weylon dark matter. The space-time is penetrated by the scalar field of the Dirac gauge function $\beta$. The latter being source of dark energy that in the very early universe create conventional matter, whereas at present the Weylian DE causes the global acceleration of the expanding universe. This classical (non quantum) 4-dimensional spatially closed cosmological model was developed in [20] from the very beginning of the universe and up to the maximum radius.

The field EQ-s are derived from the Dirac action integral with a matter term $L_{matter}$

$$I = \int [W^{\lambda\rho}W_{\lambda\rho} - \beta^2 R + \sigma\beta^2 w^\lambda w_\lambda + 2\sigma\beta w^\lambda \beta_{,\lambda} + (\sigma+6)\beta_{,\rho}\beta_{,\lambda}g^{\lambda\rho} + \quad (1)$$
$$+ 2\Lambda\beta^4 + L_{matter}]\sqrt{-g}\ d^4x$$

In (1) a comma denotes partial differentiation ($f_{,\mu} \equiv \frac{\partial f}{\partial x^\mu}$). Further $R$ is the Riemannian curvature scalar, $w^\mu$ is the Weyl connection vector, $W_{\mu\nu} \equiv w_{\mu,\nu} - w_{\nu,\mu}$ is the Weyl length curvature tensor and $\beta$ stands for the Dirac gauge function. In addition $\sigma$ is the Dirac parameter, $\Lambda$ is the cosmological constant and $L_{matter}$ stands for the Lagrangian density of matter. Varying in (1) $g_{\mu\nu}$, $w_\nu$, $\beta$ and choosing $\sigma = 0$, Dirac [14] obtained a geometrically based theory of gravitation and electromagnetism that in the Einstein gauge, $\beta = 1$, results in Einstein's general relativity theory and Maxwell's electrodynamics. Nathan Rosen [15], analyzing the W-D theory, showed that if one instead $\sigma = 0$ takes $\sigma < 0$ one gets a Proca [16] vector field, which from the quantum mechanical point of view may be treated as an ensemble of massive particles having spin 1. In the following we make use of the W-D theory enriched by Rosen's approach.

Varying in the action (1) the metric tensor, $g_{\mu\nu}$ one obtains the a' la Einstein equation

$$G^{\mu\nu} = -\frac{8\pi}{\beta^2}(T^{\mu\nu} + M^{\mu\nu}) + \frac{1}{\beta^2}(4\beta^\mu\beta^\nu - g^{\mu\nu}\beta^\lambda\beta_\lambda) + \frac{2}{\beta}(g^{\mu\nu}\beta^\lambda_{;\lambda} - \beta^\nu_{;\lambda}g^{\lambda\mu}) +$$

$$+ \frac{\sigma}{\beta^2}\left(\beta^\mu\beta^\nu - \frac{g^{\mu\nu}}{2}\beta^\lambda\beta_\lambda\right) + \frac{\sigma}{\beta}(\beta^\mu w^\nu + \beta^\nu w^\mu - g^{\mu\nu}\beta_\lambda w^\lambda) + \quad (2)$$

$$+ \sigma\left(w^\mu w^\nu - \frac{g^{\mu\nu}}{2}w^\lambda w_\lambda\right) - g^{\mu\nu}\beta^2\Lambda$$

In (2) the expression $4\pi M^{\mu\nu} \equiv \left[\frac{1}{4}g^{\mu\nu}W^{\lambda\rho}W_{\lambda\rho} - W^\mu{}_\lambda W^{\nu\lambda}\right]$ stands for the momentum-energy tensor of the $w_\nu$ – field and $T^{\mu\nu} \equiv \frac{\delta(L_{matter}\sqrt{-g})}{8\pi\sqrt{-g}\cdot\delta g_{\mu\nu}}$ for that of conventional matter. If one takes in (2) $w_\lambda = 0; \beta = 1$ one gets the EFE of G.R.

According to the standpoint presented in [20], the space-time of the universe has a chaotic Weylian microstructure but is on a large scale homogeneous and isotropic. In order to describe both, the microstructure and the global cosmic features, we regarded the Weyl vector as composed of two parts, $w_\nu = w_\nu{}_{glob} + w_\nu{}_{loc}$, with $w_\nu{}_{glob}$ being a global gradient vector, so that globally there are no vector fields. The vector $w_\nu{}_{loc.}$ represents the existing in micro-cells locally restricted chaotically oriented vector fields that create massive particles-weylons having spin 1. It was proved in [20] that the global gradient vector is given by $w_\lambda{}_{glob} = -\frac{\beta_\lambda}{\beta}$. Inserting this relation into (2) leads to the following global gravitational equation:

$$G^\nu_\mu = -\frac{8\pi}{\beta^2}T^\nu_\mu{}_{matter} + \frac{1}{\beta^2}(4\beta_\mu\beta^\nu - \delta^\nu_\mu\beta^\lambda\beta_\lambda) + \frac{2}{\beta}(\delta^\nu_\mu\beta^\lambda_{;\lambda} - \beta^\nu_{;\lambda}\delta^\lambda_\mu) - \delta^\nu_\mu\beta^2\Lambda \quad (3)$$

In the RHS of (3), $\beta = \beta(x^\mu)$; $\beta_\mu \equiv \frac{d\beta}{dx^\mu}$; $\beta^\nu \equiv g^{\nu\sigma}\beta_\sigma$ and $\beta^\nu_{;\lambda} = \beta^\nu_{,\lambda} + \beta^\sigma\{^\nu_{\sigma\lambda}\}$. The $\beta$–terms create the dark energy causing the present acceleration of the expanding universe. Further, $T^{\mu\nu}_{matter}$ stands for the momentum energy density of both, ordinary matter and dark matter–weylons created in micro-cells.

The closed homogeneous and isotropic universe was described by the F-L-R-W line element

$$ds^2 = d\tau^2 - a^2(\tau)\left[\frac{dr^2}{1-r^2} + r^2 d\vartheta^2 + r^2 \sin^2\vartheta d\varphi^2\right] \qquad (4)$$

with $a(\tau)$ being the cosmic scale parameter and $0 < r \leq 1$. From (3) with the line element (4) the cosmological equations are obtained

$$\frac{\dot{a}^2}{a^2} = \frac{8\pi}{3\beta^2}\rho_{matter} - \frac{(\dot{\beta})^2}{\beta^2} - 2\frac{\dot{a}}{a}\frac{\dot{\beta}}{\beta} + \frac{1}{3}\beta^2\Lambda - \frac{1}{a^2} \qquad (5a)$$

$$\frac{\ddot{a}}{a} = -\frac{4\pi}{\beta^2}\left(P_{matter} + \frac{1}{3}\rho_{matter}\right) + \frac{(\dot{\beta})^2}{\beta^2} - \frac{\dot{a}\dot{\beta}}{a\beta} - \frac{\ddot{\beta}}{\beta} + \frac{1}{3}\beta^2\Lambda \qquad (5b)$$

Here $\rho(\tau), P(\tau), \beta(\tau)$ are functions of $\tau$, and a dot denotes partial derivation with respect to the time $\tau$. The quantities $\rho_{matter}$ and $P_{matter}$ stand for the density and pressure of both, conventional matter and weylon DM.

In the homogeneous and isotropic W-D model [20] conventional matter is created by $\beta$–DE in the very early universe, DM is created in locally restricted micro regions by $w^\lambda$ and the $\beta$–DE causes cosmic acceleration in the late universe.

### 3. The universe with a massive cosmic body

In the previous section, a brief description of the close homogeneous isotropic universe, [20], filled with ordinary matter as well with Weylian DM and DE was given. Let us go to the universe containing a massive cosmic body. For the forthcoming discussion it is convenient to rewrite the line-element (4) in a slightly different form. With the very simple transformation, $d\tau = a(t)\cdot dt$, we get the conformal line element

$$ds^2 = a^2 dt^2 - a^2\left(\frac{dr^2}{1-r^2} + r^2 d\vartheta^2 + r^2\sin^2\vartheta d\varphi^2\right) \qquad (4a)$$

Assume a body having mass **m** is located in any point of the universe. Then, in addition to the time-dependent cosmic field discussed in [20], there is a spherically symmetric gravitational field surrounding the mass. This is the case to be considered. It will be described by the line-element

$$ds^2 = a^2 e^\nu dt^2 - a^2\left(\frac{e^\lambda dr^2}{1-r^2} + r^2 d\vartheta^2 + r^2\sin^2\vartheta d\varphi^2\right) \qquad (6)$$

with $a(t), \lambda(r)$ and $\nu(r)$. The perturbed universe is filled with conventionally matter, Weylian DM and the $\beta$–DE. These substances are now described by mass density $\tilde{\rho}(t,r)$, pressure $\tilde{P}(t,r)$ and Dirac's gauge function $\tilde{\beta}(t,r)$ all being functions of $t$ and $r$. The global field EQ. is now (3), with $\beta$ replaced by $\tilde{\beta}(t,r)$ and $T_\mu^\nu$ by $\tilde{T}_\mu^\nu$.

$$G_\mu^\nu = -\frac{8\pi}{\tilde{\beta}^2}\tilde{T}_\mu^\nu{}_{\text{matter}} + \frac{1}{\tilde{\beta}^2}\left(4\tilde{\beta}_\mu\tilde{\beta}^\nu - \delta_\mu^\nu\tilde{\beta}^\lambda\tilde{\beta}_\lambda\right) + \frac{2}{\tilde{\beta}}\left(\delta_\mu^\nu\tilde{\beta}_{;\lambda}^\lambda - \tilde{\beta}_{;\lambda}^\nu\delta_\mu^\lambda\right) - \delta_\mu^\nu\tilde{\beta}^2\Lambda \tag{3a}$$

In (6) there is a time/space separation of the metric coefficients, $g_{00} = a^2(t)\cdot e^{\nu(r)}$; $g_{11} = -a^2(t)\cdot\frac{e^{\lambda(r)}}{1-r^2}$. In accordance with this we assume $\tilde{\beta}(t,r) \equiv \overline{\beta}(t)\cdot\beta(r)$; where $\overline{\beta}(t)$ is the Dirac gauge function in the homogenous universe (cf. Sec. 2 and [20]) and $\beta(r)$ is caused by the massive body. It is also convenient to consider modified gauge functions $\tilde{b} \equiv \ln\tilde{\beta} \equiv \ln\overline{\beta} + \ln\beta = \overline{b}(t) + b(r)$, so that $\frac{\beta'}{\beta} = b'$; $\frac{\dot{\overline{\beta}}}{\overline{\beta}} = \dot{\overline{b}}$; and

$$\frac{\beta''}{\beta} = b'' + (b')^2; \quad \frac{\ddot{\overline{\beta}}}{\overline{\beta}} = \ddot{\overline{b}} + (\dot{\overline{b}})^2; \quad \frac{\dot{\tilde{\beta}}'}{\tilde{\beta}} = \dot{\overline{b}}b'.$$

Finally, from EQ-s (3a) and (6) one can obtain the gravitational equations explicitly
The $G_0^0$ EQ.

$$-3\frac{e^{-\nu}}{a^2}\left(\frac{\dot{a}}{a}\right)^2 + (1-r^2)\frac{e^{-\lambda}}{a^2}\left[\frac{\nu'-\lambda'}{r} - \frac{\nu'}{r} - \frac{2}{(1-r^2)} + \frac{1}{r^2}\right] - \frac{1}{a^2 r^2} = -\frac{8\pi}{\tilde{\beta}^2}\tilde{T}_0^0 +$$
$$+3\frac{e^{-\nu}}{a^2}\left[(\dot{\overline{b}})^2 + 2\left(\frac{\dot{a}}{a}\right)\dot{\overline{b}}\right] + (1-r^2)\frac{e^{-\lambda}}{a^2}\left[\lambda'b' + \frac{2rb'}{(1-r^2)} - \frac{4b'}{r} - 2b'' - (b')^2\right] - \tilde{\beta}^2\Lambda \tag{7}$$

the $G_1^0$ EQ.

$$\frac{e^{-\nu}}{a^2}\left[\left(\frac{\dot{a}}{a} + \dot{\overline{b}}\right)(\nu' + 2b')\right] = \frac{8\pi}{\tilde{\beta}^2}\tilde{T}_1^0 \tag{8}$$

the $G_1^1$

$$\frac{e^{-\nu}}{a^2}\left[\left(\frac{\dot{a}}{a}\right)^2 - 2\frac{\ddot{a}}{a}\right] + (1-r^2)\frac{e^{-\lambda}}{a^2}\left[\frac{\nu'-\lambda'}{r} + \frac{\lambda'}{r} + \frac{1}{r^2}\right] - \frac{1}{a^2r^2} = -\frac{8\pi}{\tilde{\beta}^2}\tilde{T}_1^1 +$$
$$+\frac{e^{-\nu}}{a^2}\left[2\ddot{\overline{b}} + (\dot{\overline{b}})^2 + 2\left(\frac{\dot{a}}{a}\right)\dot{\overline{b}}\right] - \frac{(1-r^2)\cdot e^{-\lambda}}{a^2}\left[3(b')^2 + \nu'b' + \frac{4b'}{r}\right] - \tilde{\beta}^2\Lambda \tag{9}$$

and the $G_2^2$ one.

$$G_2^2: \quad \frac{e^{-\nu}}{a^2}\left[\left(\frac{\dot{a}}{a}\right)^2 - 2\frac{\ddot{a}}{a}\right] + (1-r^2)\frac{e^{-\lambda}}{a^2}\left[\frac{\nu''}{2} + \frac{(\nu')^2}{4} - \frac{\lambda'\nu'}{4} + \frac{\nu'-\lambda'}{2r} - \frac{r\nu'}{2(1-r^2)} - \frac{1}{(1-r^2)}\right] =$$
$$= -\frac{8\pi}{\tilde{\beta}^2}\tilde{T}_2^2 + \frac{e^{-\nu}}{a^2}\left[2\ddot{\overline{b}} + (\dot{\overline{b}})^2 + 2\left(\frac{\dot{a}}{a}\right)\dot{\overline{b}}\right] + (1-r^2)\frac{e^{-\lambda}}{a^2}\left[(\lambda'-\nu')b' - 2b'' - (b')^2 + \frac{(2r^2-1)2b'}{r(1-r^2)}\right] - \tilde{\beta}^2\Lambda$$
(10).

This set of EQ-s must of course contain that of [20]. Indeed, setting in EQ-s (7-10) $v' = \lambda' = 0$. $b' = 0$ one obtains the EQ-s of the homogeneous universe.

Setting in (8) $v' = 0$; $b' = 0$ one has

$$\overline{T}_1^0 = 0 \qquad (8a)$$

Thus, as expected, there is no energy flow in the homogeneous FLRW universe.
From (7) follows

$$-\frac{3}{a^2}\left[\left(\frac{\dot{a}}{a}\right)+\dot{\overline{b}}\right]^2 - \frac{3}{a^2} = -\frac{8\pi}{\overline{\beta}^2}\overline{T}_0^0 - \overline{\beta}^2 \Lambda \qquad (7a)$$

EQ-s (9, 10) lead to

$$\frac{1}{a^2}\left[\left(\frac{\dot{a}}{a}-\dot{\overline{b}}\right)^2 - 2\left(\frac{\ddot{a}}{a}+\ddot{\overline{b}}\right) - 2(\dot{\overline{b}})^2\right] - \frac{1}{a^2} = -\frac{8\pi}{\overline{\beta}^2}\overline{T}_1^1 - \overline{\beta}^2\Lambda \qquad (9a)$$

$$\frac{1}{a^2}\left[\left(\frac{\dot{a}}{a}-\dot{\overline{b}}\right)^2 - 2\left(\frac{\ddot{a}}{a}+\ddot{\overline{b}}\right) - 2(\dot{\overline{b}})^2\right] - \frac{1}{a^2} = -\frac{8\pi}{\overline{\beta}^2}\overline{T}_2^2 - \overline{\beta}^2\Lambda \qquad (10a)$$

From (9a), (10a) follows $\overline{T}_1^1 = \overline{T}_2^2$, as expected in a homogeneous LFRW universe.

## 4. The field equations

Let us return to the universe with a cosmic body. We will represent components of the energy-momentum tensor as a product of time-depended factors and **r**-depended ones, $\widetilde{T}_\nu^\mu(t,r) = \overline{T}_\nu^\mu(t) \cdot T_\nu^\mu(r)$. The over-barred quantities defined in EQ-s (7a-10a) are given in [20], so that in (7-10) there are 4 equations for 7 functions $\lambda, v, b, T_0^0, T_0^1, T_1^1, T_2^2$. Now, by (8a) $\widetilde{T}_1^0(t,r) = \overline{T}_1^0(t) \cdot T_1^0(r) = 0$ and as the RHS of EQ. (8), , is zero, one has to fix either $v' = -2b'$; or $\frac{\dot{a}}{a} = -\dot{b}$. With the data from the previous paper [20] we see that generally $\frac{\dot{a}}{a} \neq -\dot{b}$, therefore we set

$$b' = -\frac{v'}{2} \qquad (11)$$

Condition (11) as well $\widetilde{T}_1^0 = 0$ will be adopted hereafter. Inserting (11) into (7-10) one has

$$-3\frac{e^{-v}}{a^2}\left(\frac{\dot{a}}{a}+\dot{\overline{b}}\right)^2 + \frac{(1-r^2)\cdot e^{-\lambda}}{a^2}\left[-v'' - \frac{3v'}{r} + \frac{v'-\lambda'}{r} + \frac{\lambda'v'}{2} + \right.$$
$$\left. + \frac{rv'}{1-r^2} + \frac{(v')^2}{4} - \frac{2}{1-r^2} + \frac{1}{r^2}\right] - \frac{1}{a^2 r^2} = -\frac{8\pi}{\widetilde{\beta}^2}\widetilde{T}_0^0 - \widetilde{\beta}^2\Lambda \qquad (12)$$

$$\frac{e^{-\nu}}{a^2}\left[\left(\frac{\dot{a}}{a}-\dot{b}\right)^2 - 2\left(\frac{\ddot{a}}{a}+\ddot{b}\right) - 2(\dot{b})^2\right] + \frac{(1-r^2)\cdot e^{-\lambda}}{a^2}\left[\frac{1}{r^2} - \frac{\nu'}{r} + \frac{(\nu')^2}{4}\right] -$$

$$-\frac{1}{a^2 r^2} = -\frac{8\pi}{\tilde{\beta}^2}\tilde{T}_1^1 - \tilde{\beta}^2\Lambda \qquad (13)$$

$$\frac{e^{-\nu}}{a^2}\left[\left(\frac{\dot{a}}{a}-\dot{b}\right)^2 - 2\left(\frac{\ddot{a}}{a}+\ddot{b}\right) - 2(\dot{b})^2\right] +$$

$$+ (1-r^2)\frac{e^{-\lambda}}{a^2}\left[\frac{\lambda'\nu'}{4} + \frac{\nu'-\lambda'}{2r} - \frac{\nu''}{2} + \frac{(3r^2-2)\nu'}{2r(1-r^2)} - \frac{1}{1-r^2}\right] = -\frac{8\pi}{\tilde{\beta}^2}\tilde{T}_2^2 - \tilde{\beta}^2\Lambda \qquad (14)$$

From (13, 14) one has

$$\frac{\nu''}{2} + \frac{(\nu')^2}{4} - \frac{\lambda'\nu'}{4} + \frac{\lambda'-\nu'}{2r} - \frac{r\cdot\nu'}{2(1-r^2)} + \frac{1}{r^2(1-r^2)} - \frac{e^\lambda}{r^2(1-r^2)} = \frac{8\pi\cdot a^2}{\tilde{\beta}^2(1-r^2)}\left(\tilde{T}_2^2 - \tilde{T}_1^1\right) \quad (15)$$

By $\tilde{T}_1^0 = 0$ no radial flow of matter is present. Let us consider the substance in a system of reference where it is resting, so that $\tilde{T}_2^2 = \tilde{T}_1^1$, (Cf. (9a) and (10a)). Then, EQ. (15) becomes

$$\frac{\nu''}{2} + \frac{(\nu')^2}{4} - \frac{\lambda'\nu'}{4} + \frac{\lambda'-\nu'}{2r} - \frac{r\cdot\nu'}{2(1-r^2)} + \frac{1}{r^2(1-r^2)} - \frac{e^\lambda}{r^2(1-r^2)} = 0 \qquad (16)$$

In (16) appear the metric functions $\lambda$ and $\nu$. Let us assume that $\lambda$ and $\nu$ are linked by

$$\frac{e^{\lambda+\nu}}{1-r^2} = C, \text{ with } C = Const > 0. \qquad (17)$$

Inserting this into (16) we get

$$\nu'' + (\nu')^2 - \frac{2\nu'}{r} + \frac{2}{r^2} - \frac{2Ce^{-\nu}}{r^2} = 0; \qquad (16a)$$

and turning to $y \equiv e^\nu$ one obtains

$$y'' - \frac{2}{r}y' + \frac{2}{r^2}y = \frac{2C}{r^2}. \qquad (16b)$$

From which follows

$$y \equiv e^\nu = C + Ar + Br^2, \qquad (18)$$

with $A$ and $B$ being arbitrary constants and $C > 0$ introduced in (17). From (17) and (18) follows

$$e^\lambda = \frac{C(1-r^2)}{C + Ar + Br^2}. \qquad (18a)$$

Further, $b = \ln\left[\dfrac{Const}{\sqrt{C+Ar+Br^2}}\right]$, so that taking the $Const. = \sqrt{C}$, one obtains

$$\beta(r) \equiv e^b = \dfrac{\sqrt{C}}{\sqrt{C+Ar+Br^2}}, \tag{18b}$$

It is worth noting that at $r \to 0$ one has $e^\nu \to C$; $e^\lambda \to 1$; $\beta \to 1$. At the horizon of the universe, $r=1$, one has $y \equiv e^\nu = C+A+B$; $e^\lambda = 0$; $\beta = \dfrac{\sqrt{C}}{\sqrt{C+A+B}}$

There are two interesting possibilities of interpreting $y \equiv e^\nu$ as given by (18)

1. Setting $B=0$ and $\dfrac{C}{A} = -2m$, one has $\qquad y = Ar\left(1 - \dfrac{2m}{r}\right).$ (19)

2. Denoting $\dfrac{A}{B} = -2m$; and $\dfrac{C}{B} = \kappa^2$ one rewrites

$$y \equiv e^\nu = \dfrac{C}{\kappa^2} r^2\left(1 - \dfrac{2m}{r} + \dfrac{\kappa^2}{r^2}\right); \tag{19a}$$

Adopting (19) one recognizes (in the bracket) the classic Schwarzschild metric; whereas in (19a) one has an a' la Reissner-Nordstrøm [21], [22] metric. In the following we consider the case 2. given by (19a).

Let us adopt the following meaning: ***m*** stands for the mass of the cosmic body, $\kappa$ denotes the Weyl scalar charge that creates the $\beta - $ DE around the mass ***m***. It will be assumed that $\kappa^2 < m^2$, so, one can write $\kappa = nm$ with $n < 1$.

From (18a,b) we get

$$e^\lambda = \dfrac{\kappa^2(1-r^2)}{r^2\left(1 - \dfrac{2m}{r} + \dfrac{\kappa^2}{r^2}\right)} \quad \beta(r) = \dfrac{\kappa\sqrt{C}}{r\sqrt{1 - \dfrac{2m}{r} + \dfrac{\kappa^2}{r^2}}}, \text{ and from (6) } \sqrt{|g|} = \sqrt{C}\, a^4 r^2 \sin\vartheta$$

One can rewrite (19a) as $y \equiv e^\nu = \dfrac{C}{\kappa^2}(r^2 - 2mr + \kappa^2)$, with $C > 0$, and $\kappa^2 < m^2$.

Here, as in the Reissner-Nordstrøm case, there are two horizons: the inner, Chauchi, horizon $r_{in}$ and the outer, event one, $r_{out}$.

$$r_{in} = m - \sqrt{m^2 - \kappa^2} \quad \text{and} \quad r_{out} = m + \sqrt{m^2 - \kappa^2} \tag{20}$$

Space and time change roles upon crossing the outer horizon, $r_{out}$, and change roles again upon crossing $r_{in}$. The region $r_{in} < r < r_{out}$ is a non-physical one. In the following we consider the concentration of Weyl DE around the massive body.

## 5. The DE enveloping a massive body

Dark energy (DE) does not interact with luminous and dark matter; it is affected only by the geometry. Therefore in the spherically symmetric gravitational field of the mass **m**, the $\beta$ – DE will form a ball-like concentration around **m**.
In order to evaluate the mass of this $\beta$ – ball one can turn to EQ. (7)

$$G_0^0 : -3\frac{e^{-\nu}}{a^2}\left(\frac{\dot{a}}{a}\right)^2 + (1-r^2)\frac{e^{-\lambda}}{a^2}\left[\frac{\nu'-\lambda'}{r} - \frac{\nu'}{r} - \frac{2}{(1-r^2)} + \frac{1}{r^2}\right] - \frac{1}{a^2 r^2} = -\frac{8\pi}{\tilde{\beta}^2}\tilde{T}_0^0 +$$
$$+3\frac{e^{-\nu}}{a^2}\left[\left(\dot{b}\right)^2 + 2\left(\frac{\dot{a}}{a}\right)\ddot{b}\right] + (1-r^2)\frac{e^{-\lambda}}{a^2}\left[\lambda'b' + \frac{2rb'}{(1-r^2)} - \frac{4b'}{r} - 2b'' - (b')^2\right] - \tilde{\beta}^2 \Lambda \quad (7)$$

In absent of **r**-depended $\beta$ – DE, ( $b = const.$ ), the second brackets in the RHS vanishes, so that with nonzero $b'$; $b''$ it may be regarded as the *r*-depending part of the mass-energy density of the $\beta$ – DE field

$$-8\pi\rho_\beta = (1-r^2)e^{-\lambda}\left[\lambda'b' + \frac{2rb'}{(1-r^2)} - \frac{4b'}{r} - 2b'' - (b')^2\right] \quad (21)$$

Inserting $\lambda'$; $b'$; $b''$ into (21) one obtains

$$8\pi\rho_\beta = \frac{4m}{\kappa^2 r} - 6\frac{1}{\kappa^2} + \frac{3}{\kappa^2}\cdot\frac{(r-m)^2}{(\kappa^2 - 2mr + r^2)}. \quad (21a)$$

The mass of the DE ball at *r* may be accounted by

$$M_\beta = \int_{r_0}^{r}\rho_\beta dV = \int_{r_0}^{r}4\pi\rho_\beta\sqrt{C}\frac{r^2 dr}{\sqrt{y}} = \int_{r_0}^{r}\left[\frac{2m}{\kappa^2 r} - \frac{3}{\kappa^2} + \frac{3}{2\kappa^2}\frac{(r-m)^2}{(\kappa^2 - 2mr + r^2)}\right]\frac{\kappa r^2 dr}{\sqrt{r^2 - 2mr + \kappa^2}}$$

(22)

It is convenient to evaluate the mass $M_\beta$ in two intervals separately:

A) In the region $0 < r < r_{in} = m - \sqrt{m^2 - \kappa^2}$; for extremely small *r*-s, where $y \approx \kappa = nm$.

B). Outside the event horizon for $r_{out} < r < 1$.

In the A) region for extremely small *r*-s, we obtain

$$M_\beta = \frac{m}{\kappa^2}r^2 - \frac{1}{\kappa^2}r^3 + \frac{m^2}{2\kappa^4}r^3 = \frac{r^2}{mn^2} - \frac{r^3}{m^2 n^2} + \frac{r^3}{2m^2 n^4} \quad (22a)$$

and in the first approximation for very small *r*-s

$$M_\beta \approx \frac{r^2}{mn^2} \quad (22b)$$

Substituting this into (19) one has

$$y = C\frac{M_\beta}{m}\left(1 - \frac{2m}{r} + \frac{\kappa^2}{r^2}\right); \qquad (23)$$

In this region the DE amplifies the gravitational field of **m**.

Consider the region B), outside the event horizon, $r_{out} < r < 1$. Integrating (22) one obtains (Cf. [23])

$$M_\beta = \left|\frac{3r(n^2m^2 - r^2) + 5m[r^2 - 2mr + n^2m^2]}{4nm\sqrt{r^2 - 2mr + n^2m^2}} + \frac{m(5 - 3n^2)}{4n}\cdot\ln\left[2\sqrt{r^2 - 2mr + n^2m^2} + 2r - 2m\right]\right|_{r_0}^{r} \qquad (24)$$

Let us set $r_0 \doteq r_{out} + \varepsilon \dot\approx 2m$. Then

$$M_\beta(r_0 = 2m) = \frac{m(11n^2 - 24)}{4n^2} + \frac{m}{4n}(5 - 3n^2)\ln[2m(1+n)] \qquad (24a)$$

One can show that for $r > 3r_0$, the mass $M_\beta = M_\beta(r) - M_\beta(r_0)$ is **negative**.

Let us consider a typical cosmic body, our Galaxy. One can give the following rough estimation: the mass of our Galaxy is about [27] $1.15 \times 10^{12} M_{sun}$. Taking for the present value of the cosmic scale parameter $a_{Now} \approx 1.3 \times 10^{28}$ cm we obtain in units of the radial coordinate **r** for the mass of our Galaxy $1.28 \times 10^{-11}$, so that $m_{Gal.} \ll 1$. Considering the mass of the $\beta$ – ball near the horizon ($r \approx 1$) and making use of (24, 24a) one has for the mass of the $\beta$ – dark energy ball enveloping the cosmic body **m**

$$M_{\beta Gal.} \equiv M_\beta(r = 1) - M_\beta(r = 2m) = -\frac{3}{4mn} \qquad (24b)$$

Considering the total mass of the body

$$m_{tot.} \equiv m + M_\beta = m - \frac{3}{4mn} \qquad (24c)$$

we see that it too is negative, $m_{tot.} < 0$.

By the same procedure one obtains for the model (19), at $r > 2m$ **negative** $M_\beta$ too.

Thus, the DE-ball's mass, as well the total active mass (24c) of the body, both are negative. Negative masses may be understood if we go back to the energy density (21a). For large enough values of **r** we get a very simple result, $8\pi\rho_\beta \approx -3\frac{1}{\kappa^2}$. Thus, the concentration of $\beta$ – DE around the cosmic mass create a DE-ball with negative energy-mass density and negative gravitational mass. The negative mass is universally **repulsive,** both positive-mass and negative-mass objects will be pushed away.

At the horizon of the universe ($r = 1$) one obtains

$$8\pi\rho_\beta = -\frac{3}{\kappa^2}; \text{ and } M_\beta = -\frac{3}{4\kappa} = -\frac{3}{4mn}. \qquad (25)$$

The repulsive force can be considered as a factor causing and supporting the acceleration of the expanding universe.

Going back to (9), one can consider the *r*-depending part of radial pressure created by the $\beta - \mathrm{DE}$

$$8\pi P_{\beta(r)} = (1-r^2)\cdot e^{-\lambda}\left[3(b')^2 + v'b' + \frac{4b'}{r}\right] = \frac{1}{\kappa^2}\left[\frac{(r-m)^2}{r^2-2mr+\kappa^2} - \frac{4(r-m)}{r}\right]$$

For $m \ll r \leq 1$ we have

$$8\pi P_\beta = -\frac{3}{\kappa^2}. \tag{26}$$

Thus, near the horizon of the universe the $\beta - \mathrm{DE}$ ball has a negative mass as well negative mass density (Cf. (25)) and negative pressure $8\pi P_{\beta(r)} = -\frac{3}{\kappa^2}$. It is believed that these properties are characteristic for dark energy.

## 6. Outlook and discussion

In a previous paper [20] a W-D spatially closed cyclic cosmological model was considered. The space was taken as homogeneous and isotropic, filled with luminous and dark matter as well with dark energy. These ingredients were created by geometrically features of the W-D geometry and depend on the cosmic time solely. The model [20] is in agreement with cosmological observations and predicts the acceleration of the expanding universe at present.

In the present work we consider a massive cosmic body located in the homogeneous model [20] during the dust dominated period. A mass located in the universe create a spherically symmetric gravitational field that must be considered together with the time dependent field of the previous model. The line element is taken in the form (Cf. (6)) $ds^2 = a^2 e^v dt^2 - a^2\left(\frac{e^\lambda dr^2}{1-r^2} + r^2 d\vartheta^2 + r^2 \sin^2\vartheta d\varphi^2\right)$ with $a(t)$ being the cosmic scale parameter and with $\lambda(r)$ and $v(r)$ representing the spherically symmetric deformity. The universe is filled with conventional (luminous) matter, with dark matter of weylons and with dark energy. These substances are described by $\tilde{T}^\nu_\mu(t,r)$, the momentum-energy tensor of conventional and dark matter, and by the Dirac gauge function $\tilde{\beta}(t,r)$, that creates the dark energy. All characteristics are depending on time **t** and on the radial coordinate **r**. The explicit *t*, *r* field equations (FE) are derived. It turns out, that one can achieve a *t/r* separation in the FE and one can solve the *r*-depended equations. The following solution (Cf. (19a)) is obtained $e^v = \frac{C}{\kappa^2}r^2\left(1-\frac{2m}{r}+\frac{\kappa^2}{r^2}\right); \quad e^\lambda = \frac{\kappa^2(1-r^2)}{r^2\left(1-\frac{2m}{r}+\frac{\kappa^2}{r^2}\right)}$ here **C** is a positive constant, **m** is the cosmic body's mass and $\kappa$ describes a Weylian scalar charge created in the DE environment of **m**. This charge is assumed to be smaller than the distorting mass, $\kappa^2 < m^2$. As in the Reissner-Nordstrøm case there are two horizons the inner (a Cauchi) $r_{in} = m - \sqrt{m^2 - \kappa^2}$ and the outer one

$r_{out} = m + \sqrt{m^2 - \kappa^2}$. The space is divided into three regions: $0 < r < r_{in}$, the region $r_{in} < r < r_{out}$ that is a non-physical one, and the outer $r_{out} < r \leq 1$.

In the homogeneous universe, the Weyl DE penetrated the whole space uniformly. As shown in section **5**, in the spherically symmetric gravitational field of a massive body, the $\beta$ – DE will form a ball-like concentration around the center of this body. We found that this concentration of DE has a negative mass density and a negative mass.

Negative mass in General relativity were considered by Hermann Bondi [24], and by W. B. Bonnor [25]. W. B. Bonnor and F.I. Cooperstock [26] discussed the possibility of an elementary particle (electron?) containing negative mass. Bondi [24] has considered an interesting problem of two bodies having equal absolute values of mass, but one a positive, the second a negative mass. In absent of other forces, this system would move with constant acceleration along the line from the negative mass to the positive.

In order to consider the effect of cosmic acceleration by DE let as consider the situation at the horizon of the universe, $r = 1$. Here one has (Cf. (25)) $8\pi\rho_\beta \approx -\dfrac{3}{\kappa^2}$; and $m_{tot..} \equiv m + M_\beta = m - \dfrac{3}{4mn} < 0$; the DE has also a negative radial pressure $8\pi P_\beta = -\dfrac{3}{\kappa^2}$, (CF. (26)); the latter is necessary for causing an acceleration of the expanding universe and a deceleration during the contracting phase.

The negative $m_{tot.} \equiv m + M_\beta$ is universally *repulsive,* both positive-mass and negative-mass objects placed near the horizon will be pushed away, by constant acceleration. The repulsive force is a factor causing and supporting the acceleration of the expanding universe. It is believed that negative values of $M_\beta$, $\rho_\beta$ and $P_\beta$ generally characterize DE.

In this work and in a previous one, [20], I tried, to answer the question, "What is cosmic dark energy? What is its origin, its nature?" More is unknown than is known. We know how much dark energy there is because we know how it affects the Universe's expansion. Other than that, it is a complete mystery. But it is an important mystery. It turns out that roughly 70% of the Universe is dark energy. Dark matter makes up about 25%. The rest: all "normal" matter, all stars, galaxies, intergalactic gas etc., makes less than 5% of the Universe.

My standpoint is that cosmological problems have to be considered and solved in a framework that is as close as possible to Einstein's General Theory of Relativity without adding fields, which are not geometrically based.

I would like to recall a saying of Wolfgang Pauli: "Man sollte nicht versuchen zusammen zu bringen, was der liebe Gott separat erschaffen hat."

The 4-dimensional Weyl-Dirac (W-D) theory is a minimal expansion of Einstein's theory presenting a number of geometrically based functions, so that the attempt to consider DM and DE in the W-D framework is a very natural approach.